# Dynamics of the Minority Game for Patients

Kyungsik Kim*, Seong-Min Yoon and Myung Kul Yum†

*Department of Physics, Pukyong National University, Pusan 608-737, Korea

Division of Economics, Pukyong National University, Pusan 608-737, Korea

†Department of Pediatric Cardiology, Hanyang University, Kuri 471-701, Korea

## ABSTRACT

We analyze the minority game for patients, and the results known from the minority game are applied to the patient problem consulted at the department of pediatric cardiology. We find numerically the standard deviation and the global efficiency, similar to the El Farol bar problem. After the score equation and the scaled utility are introduced, the dynamical behavior of our model is discussed for particular strategies. Our result presented will be compared with the well-known minority games.

*Corresponding author. Tel.: +82-51-620-6354; fax: +82-51-611-6357.
E-mail address: kskim@pknu.ac.kr.

Recently, the investigation of the minority game [1,2] between economists and physicists has received considerable attention as one interdisciplinary field. Until now, there has been mainly concentrated on evolutionary minority games [3-5], adaptive minority games [6,7], and multi-choice minority games [8]. Particularly, these studies can lead to a better understanding for scaling properties based on novel statistical methods and approaches of economics. Many econophysicists have considered several ways of rewarding the agent's strategies and compared the resulting behaviors of the configurations in minority game theory. Interestingly, de Almeida and Menche [8] have investigated two options rewarded in standard minority games that choose adaptive genetic algorithms, and their result is found to come close to that of standard minority games.

Furthermore, Arthor's model [1] was even more influenced to the well-known seller and buyer's model in financial markets and to the passenger problem in the metro and bus. In this paper, we apply the minority game theory to the numbers of both general patients and reserved patients who were consulted at the department of pediatric cardiology in the Korean hospital of Hanyang university for one year (from January 2002 to December 2002). For the sake of simplicity, we focus the discussion to numerically the standard deviation and the global efficiency, and we limit ourselves to the study the score equation and the scaled utility for particular strategies of our model.

In our patients problem, we will assume that the patients $N = 157$ can decide independently whether to go or not to the hospital, which the doctor can consult $x_m = (N-1)/2$ patients, i.e., the possible number of consulted patients, in a day. It is predicted that the doctor can consult them if agents become smaller then $x_m$. Patients are inductive rational agents, so that the patient does not want to go to his consulting department when the hospital is crowded. We apply $S$ different strategies to predict whether the patient will be consulted or not, based on the past consulted patients. The minority game theory, formerly introducing by Challet and Zhang [2], is applied to the dynamical behavior of our model. At each round $R$, $N$ agents (patients) choose one among two possible options, i.e., 1 and 0 (the score that the patient is decided to go or not to the hospital). The agents win (lose) if it belongs to be smaller (larger) than $x_m$, and their action behaves independently without any communication or interaction. All available information is other patint's actions, that is, the $2^m$ memory of the past $m$ rounds. There are 2 possible values for each of the $m$ memories and there are $2^m$ memories. So there exists $2^{2^m}$ different strategies, for example 256 strategies in the case of $m=3$.

A statistical quantity in the minority game is the standard deviation defined by

$$\sigma^{(p)} = [\frac{1}{R} \sum_{i=0}^{R} (N_i^{(p)} - \frac{N}{2})^2]^{1/2} \qquad (1)$$

where $R$ is the total number of rounds, $N_i^{(p)}$ is the number of agents having an arbitrary playing ($p$) in round $i$.

In our patient model, we treat with two cases as follows: There are only (1) the general patients and (2) both general and reserved patients. We will assume that one round is the consulted time for one day, and all total rounds are $R=293$, neglecting time intervals of the hospital closure. In first case, the result of standard deviations for the general patients is shown in Fig.2, which is the plot of the standard deviation $\sigma$ versus $m$ for the strategies $S=2$ and 4. It is found that the minimum value of $\sigma$ is 1.56 at $m=6$ for $S=2$, and that it approaches to the value $\sigma = 2.19$ (dot line) as $m$ goes to be larger. From the dispersion $\sigma^2$, the global efficiency $\sigma^2/N$ has a minimum value at $\alpha$ (= $2^m/N$)=0.217 for $S=2$ in Fig.3, while it obtained the same value $\alpha = 0.34$ in the El Farol bar and interacting heterogeneous model [9].

In second case, let us discuss the global efficiency of all consulted patients including reserved patients for strategy $S=2$. Fig.4 shows the standard deviation $\sigma$ as a function of the round $m$ for strategy $S=2$. Then, the minimum value $\sigma$ is 1.10 at $m=6$ and $k=20$, where $k$ is the percentage of reserved patients. As shown in Fig.5, the global efficiency has a minimum value at $m=6$ for strategy $S=2$ and $k=20$. This is an optimal situation for utilizing the hospital equipments the case of all patients including 20% reserved patients.

Next, let us $E_{S,k}(t)$ denote the score at the time step $t$ for the consulted patients per maximum patients $k$ and the strategy $S$. Since one round get the payoff if the prediction is consistent with the past recorded memory, the score equation is defined as

$$E_{S,k}(t+1) = E_{S,k}(t) + \theta [(x_t - x_m)(x_t^e - x_m)], \qquad (2)$$

where $\theta(x)$ is the Heaviside function, i.e., $\theta(x)=1$ for $x >0$ and $\theta(x)=0$ for $x <0$. The variation $x_m = (N-1)/2$ is the maximum number of consulted patients, $x_t^e$ predicted agents, and $x_t$ the number of agents chosen one particular action. As well-known, it will be rewarded with one(zero) point if a strategy is right(wrong) in Eq. (2). Fig.6 shows the score function of consulted patients for $S=2$ and the rounds $m= 2\text{-}7$ for one year. It is found that the score function becomes a larger value, as $R$ goes to be larger and m→7. Lastly we define the scaled utility $U(x_t)$ [7] as a function $x_t$ as follows:

$$U(x_t) = [(1-\theta(x_t - x_m))x_t - \theta(x_t - x_m)(N-x_t)]/x_m. \qquad (3)$$

The scaled utility at the strategy *S*=2 for one year is plotted in Fig.7. When the highest number of patints is consulted, the scaled utility has a maximum value $U(x_t = x_m$ or $x_m +1)=1$. The patients model is more efficient, as the value of the scaled utility approaches to one or the deviations from the maximum utility is small.

In conclusions, we have discussed the minority game theory for all patients included reserved patients who were consulted at the department of pediatric cardiology in the Korean hospital of Hanyang university for one year. For strategy *S*=2 and *k*=20, the global efficiency has a minimum value at the round *m*=6. In particular, we conclude that we can get the optimal situation for utilizing the hospital equipments in the case of all patients including 20% reserved patients. In future, this result will extend to other departments of Korean hospital, and it is also expected that the detail description of the minority game theory will be used to study the extension of financial analysis in Korean financial markets.

## Refernces

[1] W. B. Arthur, Amer. Econ. Review 84, 406 (1994).

[2] D. Challet and Y.-C. Zhang, Physica A246, 407 (1997); Y.-C. Zhang, Peurophys. News 29, 51 (1998).

[3] R. Savit, R. Manuca and R. Riolo, Phys. Rev. Lett. 82, 2203 (1998).

[4] D. Challet et al, Phys. Rev. Lett. 84, 1824 (2000); Phys. Rev. E60, R6271 (1999).

[5] D. Challet et al, cond-mat/0306445.

[6] L. Ein-Dor, R. Metzler, I. Kanter and W. Kinzel, Phys. Rev. E63, 066103 (2001).

[7] M. Sysi-Aho, A.Chakraborti and K. Kaski, cond-mat/0305283.

[8] J. M. L. de Almeida and J. Menche, cond-mat/030818; 0308249.

[9] D. Challet, M. Marsili and G. Ottino, cond-mat/0306445; D. Challet and M. Marsili, cond-mat/9904071.

# FIGURE CAPTIONS

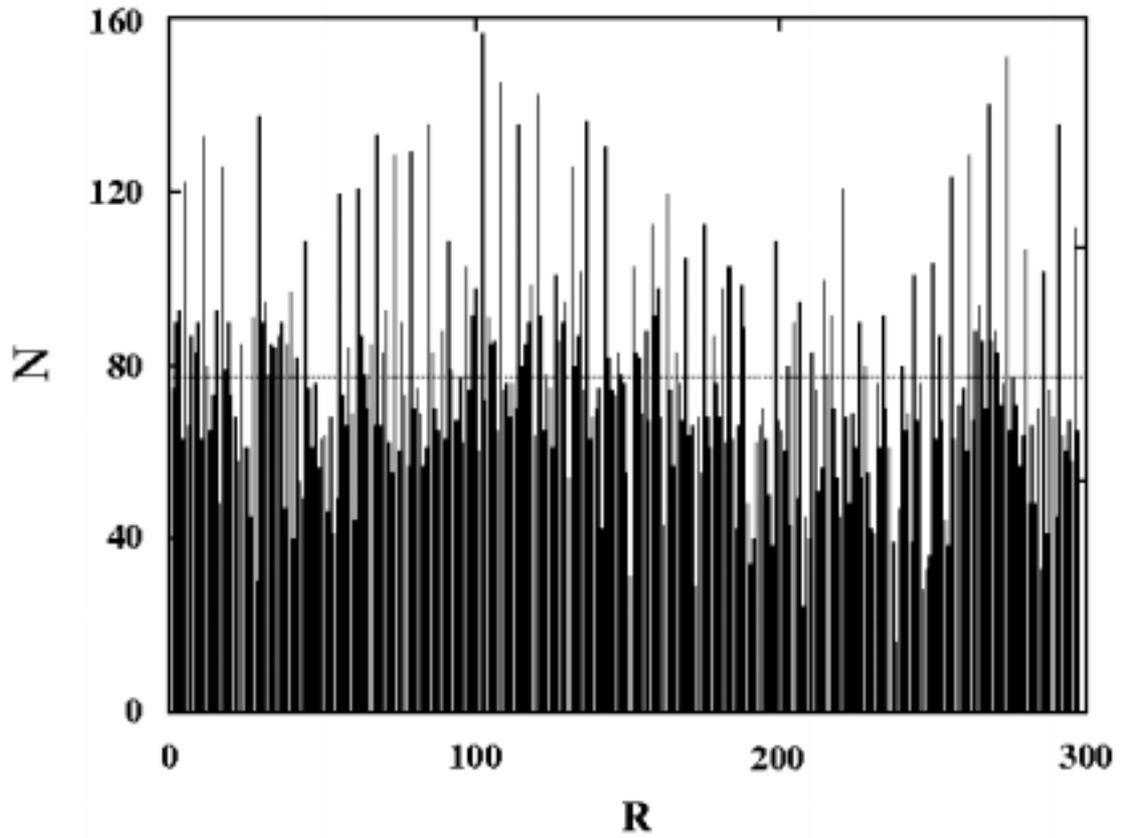

**Fig.1** The number $N$ of consulted patients for one year ($R$ =293 days), where the dot line is the value of $x_m = (N-1)/2$ and the maximum number of agents is $N = 157$.

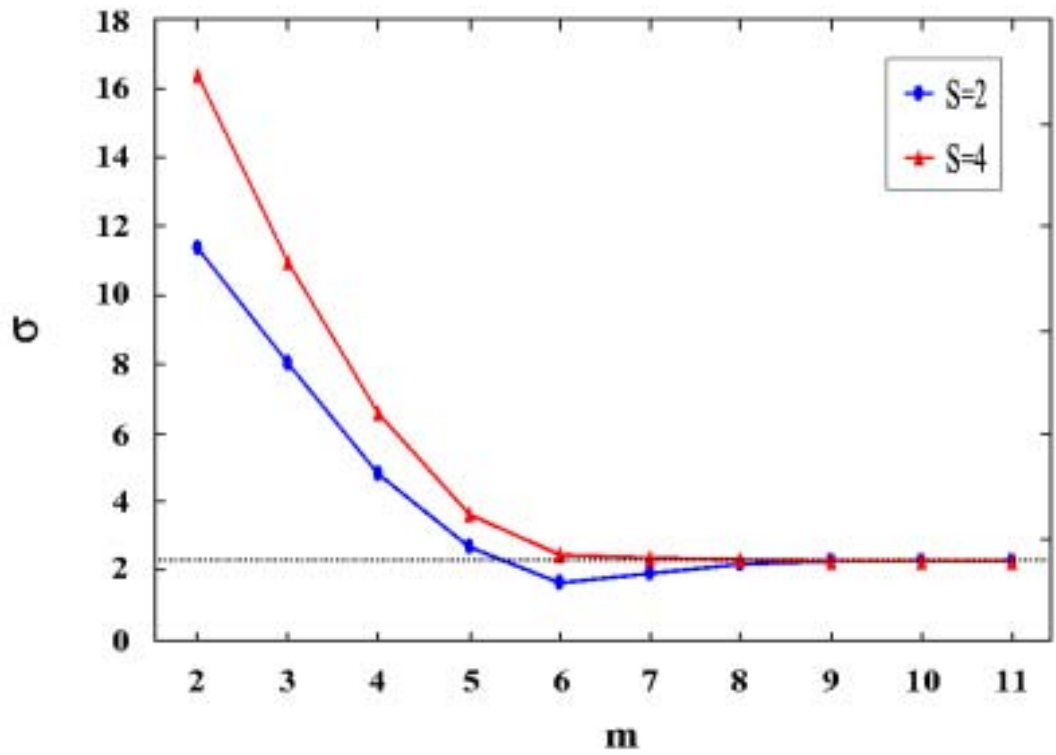

**Fig.2** Plot of the standard deviation $\sigma$ versus $m$ for the strategies S=2 and 4, where the dot line is the value of 2.19 and the minimum value of $\sigma$ is 1.56 at $m=6$ for S=2.

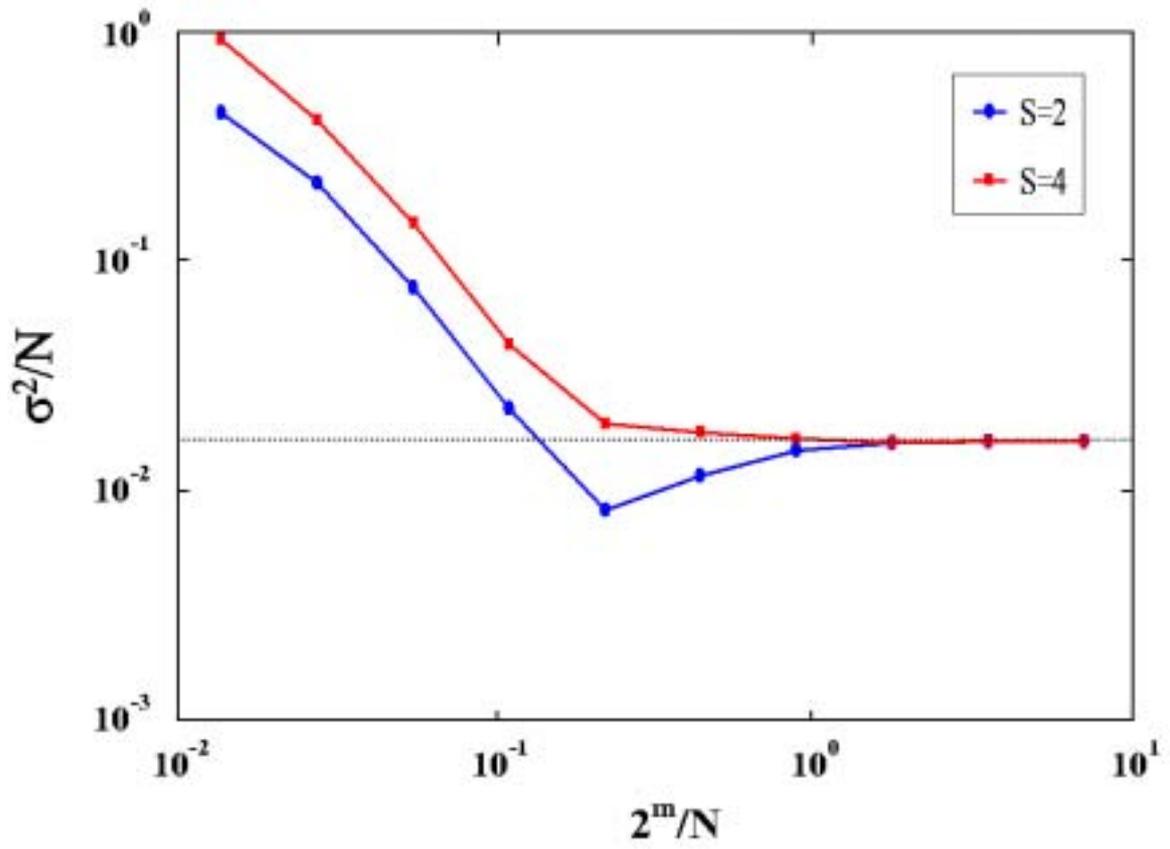

**Fig.3** Plot of the global efficiency $\sigma^2/N$ versus $2^m/N$, where the value of the dot line is -1.78 and the minimum value of $\sigma^2/N$ is –2.08 at $\alpha$ (= $2^m/N$)=0.217 for the strategy S=2.

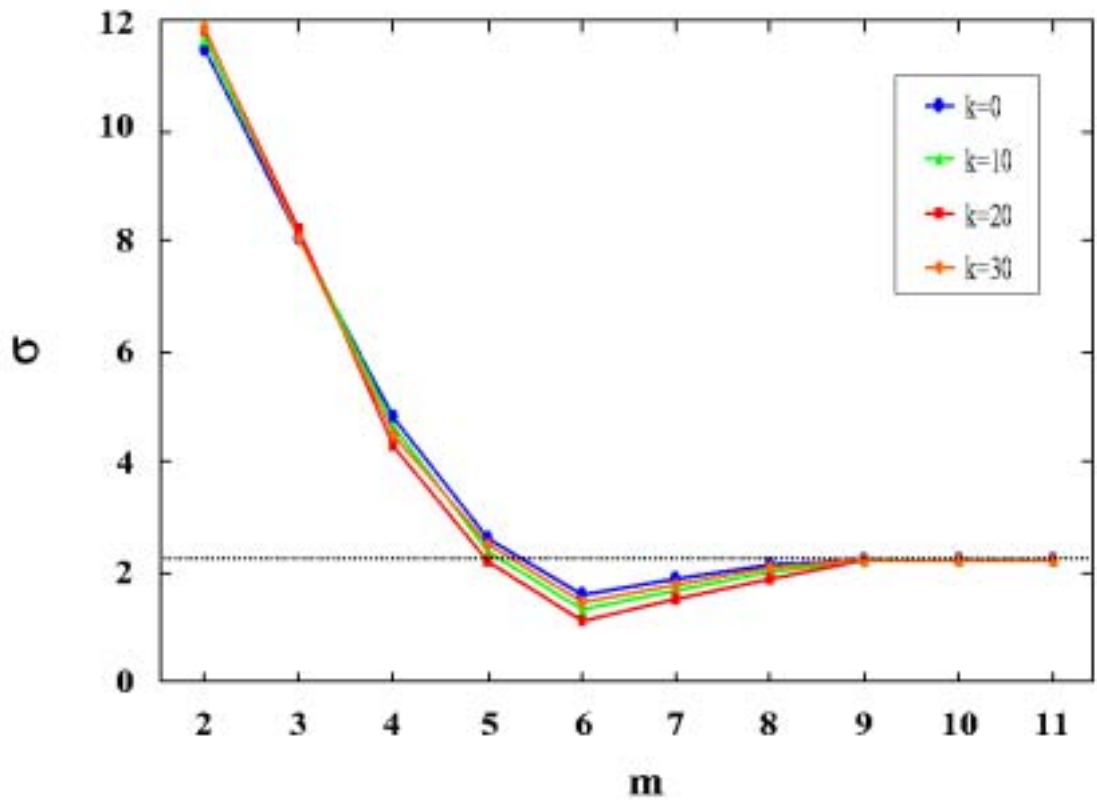

**Fig.4** Plot of the standard deviation σ versus *m* for strategy *S*=2, where the dot line is the value of 2.19 and the minimum σ is 1.10 at *m*=6 and *k*=20 ( *k* is the percentage of reserved patients).

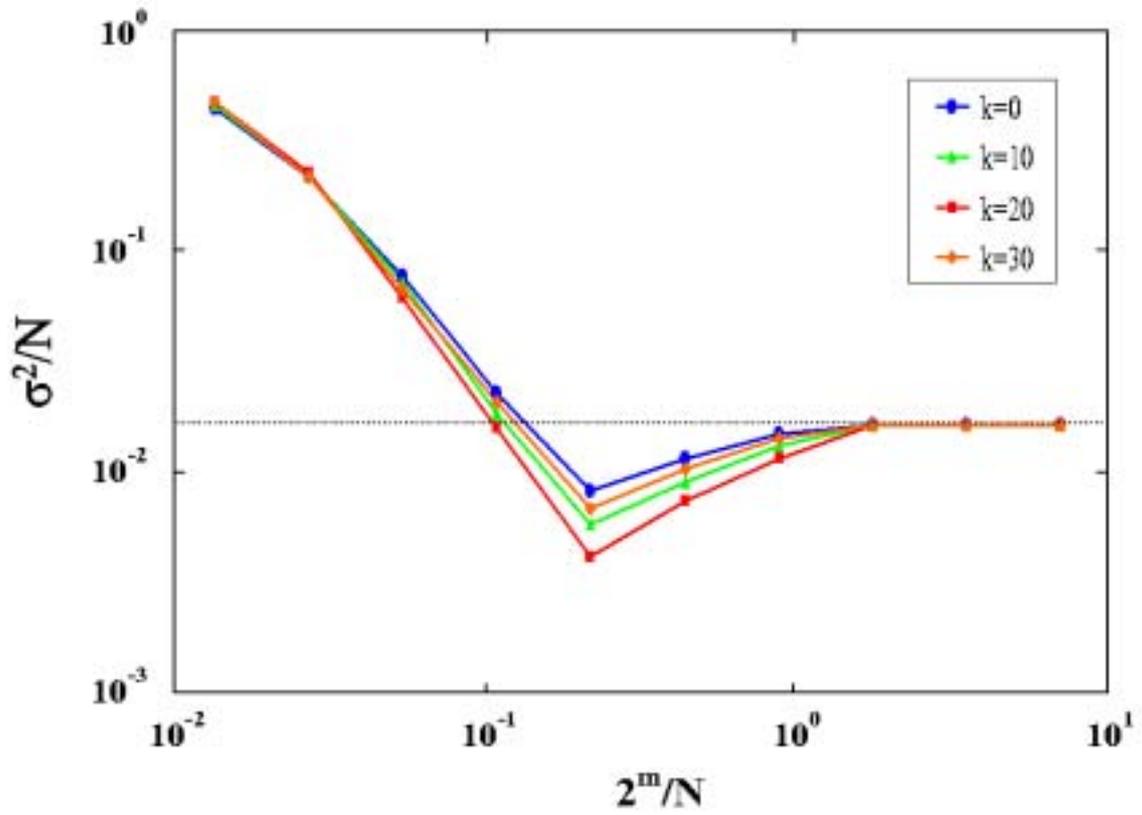

**Fig.5**  Plot of the global efficiency $\sigma^2/N$ versus $2^m/N$, where the value of the dot line is -1.78 and the minimum value of $\sigma^2/N$ is $-2.37$ at $2^m/N =0.217$ for $k=20$ and $S=2$.

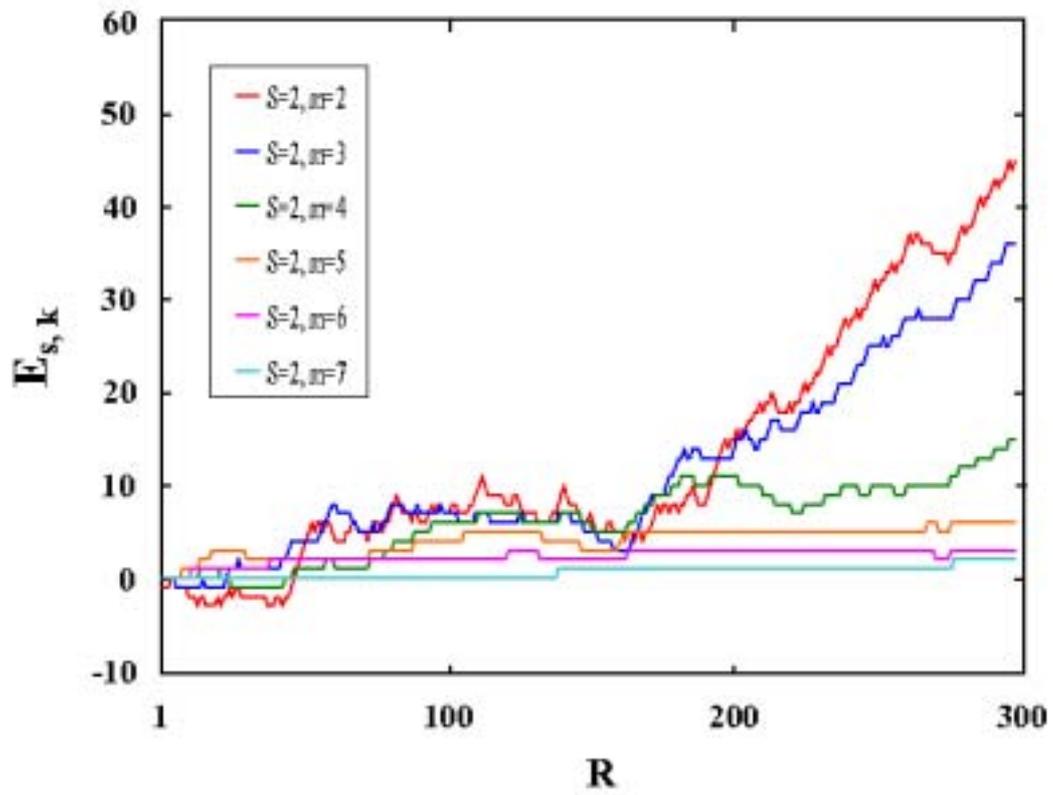

**Fig.6** The score function of consulted patients for *S*=2 and the round *m* = 2-7.

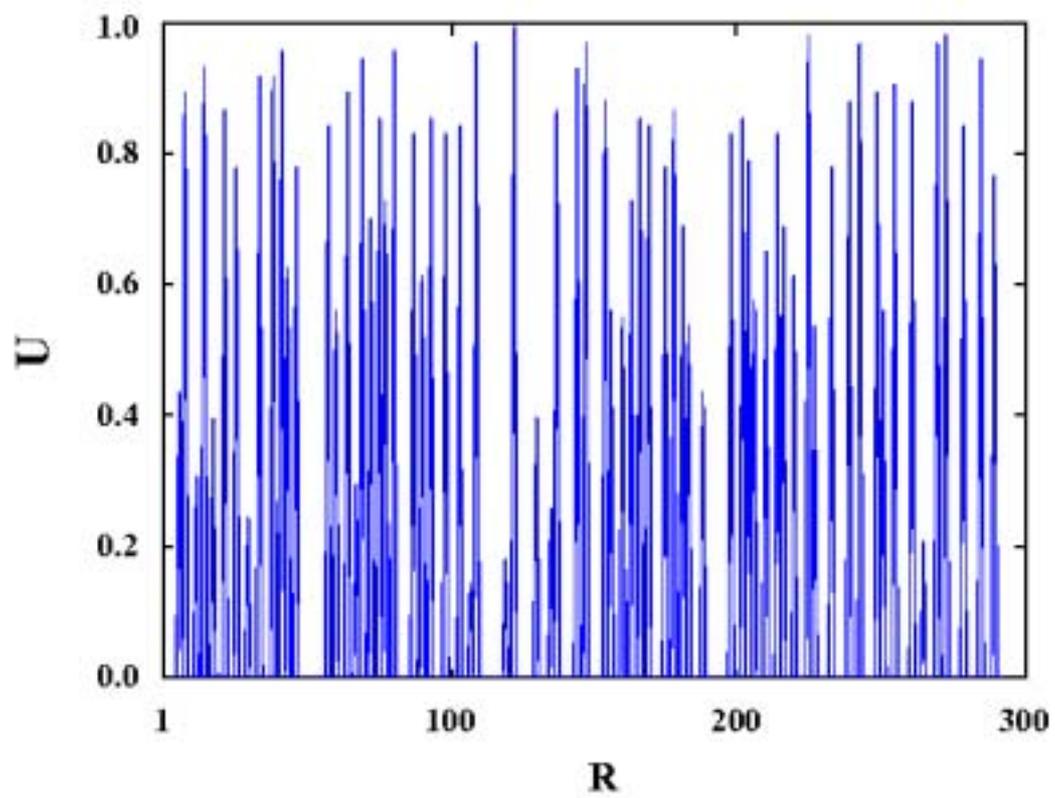

**Fig.7** The scaled utility of consulted patients at the strategy *S*=2 for one year.